\documentclass[structabstract]{aa}

\usepackage{txfonts}
\usepackage{color}
\usepackage{natbib}
\bibpunct{(}{)}{;}{a}{}{,} % to follow the A&A style
\usepackage{psfrag}
\usepackage{version}
\begin{document}

\title{The global dust modelling framework THEMIS \\[-0.2cm]
{\large (The Heterogeneous dust Evolution Model for Interstellar Solids)}}
\titlerunning{The global dust modelling framework THEMIS}
\authorrunning{A.P. Jones, M. K\"ohler, N. Ysard, M. Bocchio, L. Verstraete}

\author{A.P. Jones\inst{1} \and 
    M. K\"ohler\inst{2} \and 
    N. Ysard\inst{1}\and 
    M. Bocchio\inst{1}\and 
    L. Verstraete\inst{1}}

    \institute{Institut d'Astrophysique Spatiale, CNRS, Universit\'e Paris-Sud, Universit\'e Paris-Saclay, B\^at. 121, 91405 Orsay cedex, France \and
    School of Physics and Astronomy, Queen Mary, University of London, 327 Mile End Road, London, E1 4NS, UK \and
    \email{Anthony.Jones@ias.u-psud.fr} }

    \date{Received ? / Accepted ?}

   \abstract
{Here we introduce the interstellar dust modelling framework THEMIS (The Heterogeneous dust Evolution Model for Interstellar Solids), which takes a global view of dust and its evolution in response to the local conditions in interstellar media. This approach is built upon a core model that was developed to explain the dust extinction and emission in the diffuse interstellar medium. The model was then further developed to self-consistently include the effects of dust evolution in the transition to denser regions. The THEMIS approach is under continuous development and currently we are extending the framework to explore the implications of dust evolution in HII regions and the photon-dominated regions associated with star formation.
We provide links to the THEMIS, DustEM and DustPedia websites where more information about the model, its input data and applications can be found.}


\maketitle

%------------------------------------------------------------------
\section{Introduction}
\label{sect_intro}
%------------------------------------------------------------------

Interstellar dust has been a major topic in observational studies for more than 80 years, beginning with the early  studies of interstellar absorption. \cite{1930PASP...42..214T} studied interstellar reddening or selective absorption towards open star clusters but the effect was also, and independently, discovered by \cite{1929AN....236..249S,1931PAAS....6Q.376S}. However, Schal\'en's work on this subject in the early 1930s has largely been forgotten \citep{2014bea..book.....H}. 
Dust modelling has also had a long history and the first attempts at dust modelling appeared soon after Trumpler and Schal\'en's interstellar absorption studies. Indeed, \cite{1934MeddelAstObs_58} and \cite{1934AN....253..261S} used Gustave Mie's theory \citep{1908AnP...330..377M} to study the diffusion of starlight by interstellar matter found that  
the interstellar absorption could be reproduced by metallic particles. Sometime thereafter followed the so-called "dirty ice" model proposed by \cite{vandehulst43} and further developed by \cite{1946BAN....10..187O}. In the latter work Oort and van de Hulst determined that accretion in the interstellar medium (ISM), to a radius of 100\,nm on a time-scale $\simeq 10 ^8$\,yr, would be balanced by dust destruction in cloud-cloud collisions once every $\simeq 10^8$\,yr and that the grain life-time is $\simeq 50$ million years. Thus, dust evolution in the ISM has been an important consideration since the first dust models were developed.  
 
It was almost two decades after the dirty ice model that small graphite flakes, assumed to be formed around cool carbon stars, were proposed as an alternative dust material to explain the visible to near-IR extinction \citep{1962MNRAS.124..417H}.  Soon after silicate and graphite grains, with and without ice mantles, were shown to be consistent with the observed extinction \citep{1963MNRAS.126...99W,1970IAUS...36...42W}. With the measurement of the UV extinction and the discovery of the UV bump at $\sim 217$\,nm \citep{1965ApJ...142.1683S,1969ApJ...157L.125S} graphite became a preferred dust material and a likely UV bump-carrier  \citep{1965ApJ...142.1681S}. Soon after, it was proposed that graphite along with silicate grains, formed around oxygen-rich giant stars, are the major interstellar dust species \citep{1969Natur.223..459H}. So was born the graphite and silicate dust model, based on laboratory-measured optical constants, which after almost fifty  years remains the basis of almost all current dust models.  In the late seventies, using laboratory-measured optical constants, \cite{1977ApJ...217..425M} explored the viability of un-coated graphite, enstatite, olivine SiC, iron and magnetite as suitable interstellar dust analogues. From this they concluded that a viable model for interstellar extinction must include graphite and that it, in combination with any of the other materials, could give a good fit to the observed extinction, provided that power law grain size distributions were assumed \citep[][MRN]{1977ApJ...217..425M}. The preferred MRN dust model was for graphite grains partnered with an olivine-type silicate. Basing their work upon the MRN dust model \cite{1984ApJ...285...89D} semi-empirically tuned graphite and silicate optical properties to fit observations, resulting in "astronomical" graphite and silicate. Polycyclic aromatic hydrocarbon (PAH) molecules were later added into the mix \citep{1990A&A...237..215D,1992A&A...259..614S,1997ApJ...475..565D,2001ApJ...551..807D,2001ApJ...554..778L,2002ApJ...576..762L,2007ApJ...657..810D,2014A&A...561A..82S} but the physical basis for these models, even including the later improvements, is fundamentally little different from the MRN model. It should be noted that in all of these models the different grain populations, {\it e.g.}, graphite and olivine \citep{1977ApJ...217..425M}, "astronomical" graphite and silicate \citep{1984ApJ...285...89D} or "astronomical" graphite, silicate and PAH \citep{1992A&A...259..614S,1997ApJ...475..565D,2001ApJ...551..807D,2001ApJ...554..778L,2002ApJ...576..762L,2007ApJ...657..810D,2014A&A...561A..82S} are considered to reside in distinct and separate dust populations. Some later dust models followed this same basic approach but abandoned "astronomical" graphite in favour of physically more-realistic amorphous carbons \citep[{\it e.g.},][]{2004ApJS..152..211Z,2011A&A...525A.103C,2011A&A...536A..88G}.  However, most of these models also suffer the same limitations as all MRN-based models, {\it i.e.}, ``The particles responsible for the 2200\,\AA\ hump and the FUV extinction constitute two independent populations, and the MRN graphite plus silicate model violates this condition.'' \citep{1983ApJ...272..563G} and ``The ultraviolet extinction by the particles responsible for the visual extinction is gray and therefore these particles constitute an entirely separate population from the hump particles or the FUV particles.'' \citep{1983ApJ...272..563G}. In contrast, the \cite{1990A&A...237..215D} dust model was built to be consistent with \cite{1983ApJ...272..563G} requirements and was something of a departure from previous models in that it introduced PAHs, assumed that the small grains are made of an amorphous carbon material and adopted big silicate grains with a ``dark refractory mantle'' of hydrocarbon composition. 

Given the turbulent nature of the ISM and the observation that the depletions of the dust-forming elements vary significantly and differentially \citep[{\it e.g.},][]{1952ApJ...115..227R,1994ApJ...424..748C,1996ARA&A..34..279S,2000JGR...10510257J,2002ApJ...579..304W,2009ApJ...700.1299J,2012ApJ...760...36P} it is difficult to see how the different dust materials could remain distinct. The cycling of gas and dust through different ISM phases naturally requires that the dust materials become mixed into inhomogeneous assemblages. This mixing could be in the form of graphite and silicate core ice-mantled grains \citep{1963MNRAS.126...99W} or refractory core/mantle particles, along the lines of those originally proposed by \cite{1986Ap&SS.128...17G}, in porous particles or mixed-material aggregates \citep[{\it e.g.},][]{1988MNRAS.234..209J,1989ApJ...341..808M}. Indeed, physically-viable interstellar dust models based on silicate grains with "organic" or (hydrogenated) amorphous carbon mantles have been around for more than 30 years \citep[{\it e.g.},][]{1986Ap&SS.128...17G,1987MNRAS.229..203D,1987MNRAS.229..213J,1989MNRAS.236..709D,1990QJRAS..31..567J,1990MNRAS.247..305J,1991MNRAS.248..439S,1997A&A...323..566L,2008MNRAS.384..591I,2010MNRAS.408..535C,2011MNRAS.410.1932Z,2013A&A...558A..62J,2014ApJ...785...41C,2014ApJ...788..100C,2014A&A...565L...9K,Faraday_Disc_paper_2014}. As per all dust models, variations in the grain size distribution can be used to explain the observed variations in dust emission and extinction. However, a major advantage of the core/mantle models is that they can, in addition and quite naturally, account for dust variations through environmentally-driven changes in the carbonaceous mantle composition and/or its depth \citep[{\it e.g.},][]{1987MNRAS.229..213J,1989MNRAS.236..709D,1990QJRAS..31..567J,1997A&A...323..566L,2010MNRAS.408..535C,2011MNRAS.410.1932Z, 2013A&A...558A..62J,2014A&A...565L...9K, 2014ApJ...785...41C,2014ApJ...788..100C,2015A&A...577A.110Y,2016A&A...588A..43J,2016A&A...588A..44Y}. 

The long-standing and widely-adopted idea of core/mantle or CM interstellar grains was recently given a new treatment \citep{2013A&A...558A..62J,2014A&A...565L...9K}. The result of this is the THEMIS dust modelling framework, which provides a core/mantle model for dust in the diffuse ISM and the evolution of the dust properties in response to their local environment \citep{2013A&A...558A..62J,2014A&A...565L...9K,Faraday_Disc_paper_2014,2014A&A...570A..32B,2015A&A...577A.110Y,2016A&A...588A..43J,2016A&A...588A..44Y}. The underlying principle of the THEMIS modelling framework is the supposition that interstellar dust is not the same everywhere but that it evolves within a given region of the interstellar medium (ISM) as it reacts to and interacts with its local environment. Indeed, variations in the dust properties from one region to another have long been interpreted as due to dust evolution \citep[{\it e.g.},][]{1974A&A....37...17L,1980ApJ...235...63J,1983ApJ...272..551A,1992ApJ...386..562W,1993A&A...280..617O,2001ApJ...547..872W,2003A&A...398..551S}. For example, photon, ion and electron irradiation can induce changes in the dust chemical composition and structure \citep[{\it e.g.},][]{2001A&A...368L..38D,2004A&A...420..233D}, hydrogenation and accretion can drive changes in the grain chemical composition \citep[{\it e.g.},][]{1986ApJ...305..817H,1990MNRAS.243..570S,1991MNRAS.248..439S,2008ApJ...682L.101M,2010ApJ...718..867M,2012A&A...540A...1J} 
and accretion/coagulation will change the grain structure \citep[{\it e.g.},][]{2011A&A...528A..96K,2012A&A...548A..61K,Faraday_Disc_paper_2014,2015A&A...579A..15K,2016A&A...588A..43J}. 
All of these processes directly affect the dust optical properties ({\it i.e.}, the particle absorption and scattering cross-sections) as the composition, structure and shape of the grains evolve as they transit from one region to another.  For example, in the tenuous ISM the outer carbonaceous layers of the grains, be they carbon grains or the mantles on other grains, will be H-poor and aromatic rich due to UV photolysis by stellar FUV/EUV photons \citep[{\it e.g.},][]{1986ApJ...305..817H,1990MNRAS.243..570S,1991MNRAS.248..439S,2012A&A...540A...1J,2012A&A...540A...2J,2012A&A...542A..98J,2013A&A...558A..62J,2016RSOS....360221J}   but in denser regions accreted hydrocarbon mantles are likely H-rich and aromatic poor \citep{Faraday_Disc_paper_2014,2016A&A...588A..43J,2016A&A...588A..44Y}. 

In our dust modelling studies we consider the dust properties as they evolve in response to the physical conditions in their local environment, {\it e.g.},  density, radiation field (intensity and hardness as effected by extinction)  and kinematics/dynamics. This evolution drives the dust properties (structure and composition) towards a state that is assumed to be in equilibrium with the given local conditions. For example, in diffuse and dense clouds, where the dust seemingly resides undisturbed for millions of years, the typical dust processing time-scales are short with respect to the dynamical time-scales and the dust is generally in equilibrium with the local conditions in these regions. However, in photon-dominated regions (PDRs) and shocked regions this is no longer the case because the dynamical time-scales are significantly shorter and the dust is  often out of equilibrium and constantly evolving in response to the local physical conditions. 

Here and from now on we will refer to our evolutionary dust modelling approach, and all future developments and extensions of it, under the umbrella acronym THEMIS (The Heterogeneous dust Evolution Model for Interstellar Solids).\footnote{In previous publications THEMIS was the acronym for The Heterogeneous dust Evolution Model at the Institut d'astrophysique Spatiale.}

%------------------------------------------------------------------
\section{The THEMIS dust modelling framework}
\label{sect_themis}
%------------------------------------------------------------------

The central tenet of THEMIS is that interstellar dust is not the same everywhere but that it reacts to its local environment. The core dust model central to the THEMIS approach (see Section~\ref{sect_model}) is built, as much as has been possible, upon the foundations of the laboratory-measured properties of physically-reasonable interstellar dust analogue materials, {\it i.e.}, amorphous olivine-type and pyroxene-type silicates with iron and iron sulphide nano-inclusions, a-Sil$_{\rm Fe,FeS}$, and the extensive family of hydrogenated amorphous carbon materials, a-C(:H) \citep{2013A&A...558A..62J,2014A&A...565L...9K,2015IAUGA..2256693J}. In constructing and developing the THEMIS framework we consider that the physical properties of a-Sil$_{\rm Fe,FeS}$ and, in particular, of a-C(:H) materials and their structural juxtapositions evolve in a coherent and self-consistent manner within a given region of the interstellar medium (ISM). For example, irradiation can induce changes in the dust chemical composition and structure, and changes in structure can also be driven by accretion and coagulation. All of these processes  directly affect the dust optical properties (absorption and scattering cross-sections), which are the key to understanding the nature of dust. In our modelling we take into account the shape of the particles and the evolution of the shape distribution along a sequence of increasingly complex particle shapes (see Section \ref{sect_dense}).  Shape effects are particularly important in the case of coagulated particles and grains with incomplete and/or non-uniform  mantles \citep[{\it e.g.},][]{2012A&A...548A..61K}. For complex grain shapes we calculate the optical properties using DDSCAT \citep{2000ascl.soft08001D}, as described elsewhere \citep{2011A&A...528A..96K,2012A&A...548A..61K,2015A&A...579A..15K}.  We have not yet considered the polarising properties of such complex grain structures. 

In the following sub-sections we highlight the most important aspects of THEMIS and how this framework approach will be developed in the future.

%------------------------------------------------------------------
\subsection{Dust structural and optical properties}
\label{sect_props}
%------------------------------------------------------------------

At the core of any model for interstellar dust are the input optical constants, the complex indices of refraction ($m = n + ik$), which are used to derive the dust cross-sections (optical properties), for extinction ($i = $ ext), absorption ($i = $ abs) and scattering ($i = $ sca), 
\begin{equation}
C_i(a,\phi,\xi,\lambda) = \pi a^2 \times Q_i(a,\phi,\xi,\lambda), 
\label{eq_sigma}
\end{equation}
which are a function of the particle radius ($a$), material ($\phi$), grain structure ($\xi$) and the wavelength ($\lambda$). 
We have as much as possible based our dust model on the best available, laboratory-measured or laboratory-derived properties of analogues of interstellar dust materials:  amorphous silicates \citep{1996ApJS..105..401S},  
hydrogenated amorphous carbons \citep{1984JAP....55..764S,1984ApJ...287..694D,1991ApJ...377..526R,1995ApJS..100..149M,1996MNRAS.282.1321Z},  iron \citep{1983ApOpt..22.1099O,1985ApOpt..24.4493O,1988ApOpt..27.1203O}, iron sulphide \citep{1994ApJ...421..615P} and water ice \citep{Warren_1984_AO}. Where the available data do not cover the required (wavelength) range we have extrapolated those data in as physically-meaningful way as possible. Further, where suitable data were not available we have constructed new data `ground-up' using fundamental principles and have calibrated these data on the available laboratory measurements \citep{2012A&A...540A...1J,2012A&A...540A...2J,2012A&A...542A..98J,2014A&A...565L...9K}. 

%------------------------------------------------------------------
\subsubsection{Amorphous silicates, a-Sil}
\label{sect_aSil}
%------------------------------------------------------------------

In the case of amorphous silicates, a-Sil, there is a wealth of laboratory data and it can be difficult to determine which is the most appropriate analogue material data to use to model interstellar silicates. Hence, in our choice, we were motivated by 
experimental data, astronomical observations, direct interstellar dust analyses and interstellar depletion studies of, principally, Mg, Si, Fe and S.  Our final choice of mixed amorphous silicates with metallic iron and iron sulphide nano-inclusions was motivated by three major lines of evidence. A fourth piece of evidence subsequently came to light and thus adds weight to our earlier assumptions. 

Firstly,  observations of silicate dust in the ISM towards the Galactic Centre indicate that they are a mix of amorphous olivine-type and pyroxene-type silicates  \cite[{\it e.g.},][]{2000ApJ...537..749C}.

Secondly, the work of \cite{2006A&A...448L...1D} on the annealing of iron-bearing amorphous silicate shows that in the presence of carbon the iron is reduced to metal nano-inclusions within an Mg-rich amorphous silicate matrix. This structure is much akin to the Glass with Embedded Metal and Sulphide (GEMS) component of interplanetary dust particles \citep[{\it e.g.},][]{1994Sci...265..925B}

Thirdly, X-ray observational data indicate that even though iron is present in interstellar dust it is not in Fe-rich silicates but in the form of metallic iron inclusions within Mg-rich silicates \cite[{\it e.g.},][]{2005A&A...444..187C,2012A&A...539A..32C,2011ApJ...738...78X}

Fourthly, and subsequent to the publication of the original \cite{2013A&A...558A..62J} diffuse ISM dust model, seven particles of interstellar origin collected and analysed by the Stardust mission exhibit amorphous and  crystalline grains and multiple iron-bearing phases, including metallic iron and iron sulphide \citep{2014Sci...345..786W}. 

For interstellar amorphous silicate dust, a-Sil, we use a 1:1 by mass mix of amorphous forsterite-type and enstatite-type silicates \citep[with the complex refractive indices $n$ and $k$ taken from][]{1996ApJS..105..401S} with iron incorporated in the form of metallic iron iron \citep[$n$ and $k$ from][]{1983ApOpt..22.1099O,1985ApOpt..24.4493O,1988ApOpt..27.1203O} and  iron sulphide \citep[$n$ and $k$ from][]{1994ApJ...421..615P} nano-particle inclusions occupying 7\% and 3\% of the grain volume, respectively  \citep[a-Sil$_{\rm Fe,FeS}$,][]{2013A&A...558A..62J,2014A&A...565L...9K}.\footnote{The complex refractive indices were calculated separately for each a-Sil material mixed with Fe,FeS nano-inclusions using the Garnett effective medium theory \citep[EMT,][]{1904RSPTA.203..385G}.}

%------------------------------------------------------------------
\subsubsection{Amorphous silicate mass densities}
\label{sect_rho_aSil}
%------------------------------------------------------------------

In our dust modelling we assume an amorphous silicate material density of 2.5\,g\,cm$^{-3}$, which has probably never before been justified. The assumption of a lower density, than for bulk silicate materials, is adopted because, firstly, amorphous materials are less dense than their crystalline cousins and, secondly, sub-$\mu$m particles will not have the same density as their larger siblings or their parental solids because of surface effects. 

For the crystalline silicate mineral equivalents of the amorphous silicate materials that we use, {\it i.e.}, enstatite-type pyroxene (MgSiO$_3$) and forsterite-type olivine (Mg$_2$SiO$_4$),\footnote{N.B. The amorphous equivalents of the crystalline silicate should be referred to as of {\em -type} because these specific mineral names only refer to crystalline materials.} the bulk material densities are 3.2\,g\,cm$^{-3}$ and 3.3\,g\,cm$^{-3}$, respectively. 
The assumed a-Sil density is therefore $\simeq 20$\% lower than the equivalent crystalline silicate material density. 

While is it hard to find direct measurements of the  densities of enstatite-type and forsterite-type amorphous silicate nano-particles, it has been shown that silica (SiO$_2$) nano-particles ($a \simeq 30 - 200$\,nm) have densities of $\sim 1.9$\,g\,cm$^{-3}$ \citep{2014IAC_silica_nps} compared to a bulk material density of $2.2-2.7$\,g\,cm$^{-3}$. The density of silica nano-particles is therefore reduced by $\simeq 14-30$\% with respect to that of the parental solid. 
Thus, and by analogy with silica, a density reduction of $\simeq 20$\% for sub-$\mu$m amorphous silicate particles would appear to be supported by laboratory measurements \citep{2014IAC_silica_nps}.

%---------- TABLE
\begin{table*} 
\caption{The THEMIS model elemental abundance requirements (in ppm). Variations in the diffuse ISM dust emission and extinction \citep{2015A&A...577A.110Y} can be explained with a range of a-C mantle depths (thin$-${\it thick}, see text), which do not significantly affect the total carbon abundance.}
\begin{center}
\begin{tabular}{lccccccccccc}
\hline
\hline
       &      &     &     &     &     &     &    &    &    &       &        \\[-0.25cm]
   Dust    &   C   &  C   &   C  &     &     &      &   Fe   &  Fe  &    &  density  &       \\ 
   component   &   (a-C:H)   &  (a-C)   &   (total)  &  O   &  Mg   &  Si   &  (metal)    &  (FeS)   &  S  &  (g/cm$^3$)  &   M$_{\rm d}$/M$_{\rm H}$    \\ 
  \hline
      &      &     &     &     &     &     &    &    &    &       &       \\[-0.25cm] 
  nano a-C  &   13$-${\it 4}   &  130$-${\it 140}   &   143$-${\it 144}  &  ---   &  ---   &   ---  &  ---    &  ---   &  ---  &  1.60  &  $1.7 \times 10^{-3}$  \\ 
  large a-C:H/a-C   &   45$-${\it 39}   &  5$-${\it 13}   &   50$-${\it 52}  &   ---  &   ---  &   ---  &  ---     &  ---  &  ---  &  1.51$-${\it 1.57}  &  (6.0$-${\it 6.3})\,$\times 10^{-4}$  \\ 
  large a-Sil/a-C   &  ---   &   13$-${\it 22}   &  13$-${\it 22}   &   110  &  45   &  32   &  16    &   3   & 3   &  2.19$-${\it 1.94}  &   (5.1$-${\it 4.5})\,$\times 10^{-3}$ \\ 
      \hline
\end{tabular}     
\end{center}    
\label{table_abundances}
\end{table*}
%----------

%------------------------------------------------------------------
\subsubsection{Hydrogenated amorphous carbons, a-C(:H)}
\label{sect_aCH}
%------------------------------------------------------------------

Finding definitive laboratory data sets for physically-reasonable analogues of interstellar carbonaceous materials is difficult because of the wide range in properties of these materials \cite[{\it e.g.},][]{1986AdPhy..35..317R,1988PMagL..57..143R,1987PhRvB..35.2946R}. Graphite has traditionally been used in many dust models but, while there are rather rare pre-solar graphite grains in primitive meteorites, there appears to be no evidence to support it as the most abundant interstellar carbon dust material \citep[{\it e.g.},][]{2004A&A...423L..33D,2005A&A...432..895D,2008A&A...492..127S,2011A&A...525A.103C}. However, amorphous hydrocarbons do appear to be an important component of the dust in the Milky Way 
\citep[{\it e.g.},][]{2002ApJS..138...75P} and also in galaxies \citep[{\it e.g.},][]{2004A&A...423..549D}. 
These materials are consistent with the observed infrared absorption bands \citep[{\it e.g.},][]{2002ApJS..138...75P,2004A&A...423L..33D,2007A&A...476.1235D} and show temperature-dependent luminescence  \citep[{\it e.g.},][]{1986AdPhy..35..317R} consistent with the observed interstellar luminescence in the red, the so-called extended red emission \citep[ERE,][]{1997ApJ...482..866D,2001ApJ...553..575D,2005A&A...432..895D,2010A&A...519A..39G} and with the likely irradiation effects and the associated aliphatic to aromatic transformation in the ISM \citep[{\it e.g.},][]{1990MNRAS.247..305J,2004A&A...423..549D,2004A&A...423L..33D,2008A&A...490..665P,2008ApJ...682L.101M,2011A&A...529A.146G}, in circumstellar regions \citep[{\it e.g.},][]{2003ApJ...589..419G,2007ApJ...662..389G,2007ApJ...664.1144S,2008A&A...490..665P}, in interplanetary dust particles \citep[IDPs,][]{2006A&A...459..147M} and solar system organics \citep{2011A&A...533A..98D}. Thus, the most physically-realistic carbonaceous grain materials, albeit inherently rather complex, would appear to be the extensive family of (hydrogenated) amorphous carbons, a-C(:H).\footnote{These semiconducting materials span a wide compositional range; from H-poor, aromatic-rich a-C with $\lesssim 15$ at. \% H and narrow band gap ($E_{\rm g} < 1$\,eV) to H-rich, aliphatic-rich a-C:H with $\sim15-60$ at. \% H and wide band gap ($E_{\rm g} \geq 1$\,eV).} Hydrogenated amorphous carbon materials, a-C(:H), are macroscopically-structured, contiguous network, solid-state materials comprised of only carbon and hydrogen atoms. The properties of a-C(:H) materials have been well-studied within both the physics and astrophysics communities \citep[{\it e.g.},][]{1979PhRvL..42.1151P,1980JNS...42...87D,1983JNCS...57..355T,1986AdPhy..35..317R,1987PhRvB..35.2946R,1988JVST....6.1778A,1988Sci...241..913A,1988PMagL..57..143R,1990JAP....67.1007T,1991PSSC..21..199R,1995ApJS..100..149M,1996ApJ...464L.191M,2000PhRvB..6114095F,2001PSSAR.186.1521R,2002MatSciEng..37..129R,2003ApJ...587..727M,2004PhilTransRSocLondA..362.2477F,2007DiamondaRM...16.1813K,2007Carbon.45.1542L,2008ApJ...682L.101M,2011A&A...528A..56G,2012A&A...540A...1J,2012A&A...540A...2J,2012A&A...542A..98J}. 

There is a large database of laboratory-measured \cite[{\it e.g.},][]{1984JAP....55..764S,1984ApJ...287..694D,1995ApJS..100..149M,2016A&A...586A.106G}\footnote{ See also the Jena group's ``Databases of Dust Optical Properties'' http://www.astro.uni-jena.de/Laboratory/Database/databases.html} and post-processed  \cite[{\it e.g.},][]{1991ApJ...377..526R,1996MNRAS.282.1321Z}\footnote{N.B. These two post-processed data sets, derived from similar laboratory data, arrive at divergent solutions for the optical properties of their amorphous carbons at far- to extreme-UV wavelengths.} optical property data for a-C(:H) materials but it is far from homogeneous in terms of composition/structure, synthesis methods and wavelength coverage. Hence, until such time as the gaps have been filled we rely upon a recently-developed set of optical properties for a large part of the a-C(:H) family parameter space \citep{2012A&A...540A...1J,2012A&A...540A...2J,2012A&A...542A..98J}. These optical properties were built ground-up using random covalent network (RCN) models \citep{1979PhRvL..42.1151P,1980JNS...42...87D,1983JNCS...57..355T,1988JVST....6.1778A} and extensions thereof \citep[extended RCN (eRCN) and Defective Graphite (DG) models,][]{1990JAP....67.1007T,1990MNRAS.247..305J,2012A&A...540A...1J} to construct a solid-state framework method to derive their structure and thence their complex indices of refraction. This approach determines the composition- and size-dependent optical properties of a-C(:H) materials \citep[the optEC$_{\rm (s)}$ and optEC$_{\rm (s)}$(a) datasets,][]{2012A&A...540A...1J,2012A&A...540A...2J,2012A&A...542A..98J} based upon eRCN and DG modelling methods \citep{1990MNRAS.247..305J,2012A&A...540A...1J}. Using this approach the likely evolution of hydrocarbon solids in the ISM was elucidated and explored in detail, even down to molecular dimensions, by \cite{2012A&A...540A...1J,2012A&A...540A...2J,2012A&A...542A..98J}. THEMIS builds upon this foundation, which was used to construct a diffuse ISM dust model \citep{2013A&A...558A..62J,2014A&A...565L...9K}, using the optEC$_{\rm (s)}(a)$ optical property data\footnote{Available at: \\ http://cdsarc.u-strasbg.fr/viz-bin/qcat?J/A+A/545/C2 \\ and http://cdsarc.u-strasbg.fr/viz-bin/qcat?J/A+A/545/C3.} to predict and explore the observed interstellar dust extinction and emission properties (see following section).

%------------------------------------------------------------------
\subsection{The diffuse ISM dust model}
\label{sect_model}
%------------------------------------------------------------------

THEMIS makes the fundamental assumption that the silicate and carbonaceous dust populations cannot be completely segregated in the ISM and that amorphous silicate grains are mixed with a carbonaceous dust component \cite[{\it e.g.},][]{1989ApJ...341..808M,1990A&A...237..215D}, most likely in core/mantle (CM) structures \cite[{\it e.g.},][]{1986Ap&SS.128...17G,1987MNRAS.229..203D,1987MNRAS.229..213J,1989MNRAS.236..709D,1990QJRAS..31..567J,1997A&A...323..566L}. The core of the THEMIS approach is its diffuse ISM dust model \citep{2013A&A...558A..62J,2014A&A...565L...9K,2015A&A...577A.110Y}, which comprises the following grain and core/mantle (CM) grain structures and compositions: 
\begin{itemize}
\item  a power-law distribution of a-C nano-particles ($a \lesssim 20$\,nm) with strongly size-dependent optical properties,  
\item  a log-normal distribution ($a \simeq 10-3000$\,nm, $a_{\rm peak} \simeq 160$\,nm) of large a-C(:H) CM grains with UV photolysed a-C surface layers (mantle depth $= 7.5 - 20$\,nm) surrounding a-C:H cores, and 
\item  a log-normal distribution ($a \simeq 10-3000$\,nm, $a_{\rm peak} \simeq 140$\,nm) of large a-Sil$_{\rm Fe,FeS}$ grains with a-C mantles (depth $= 5 - 10$\,nm) formed by carbon accretion and/or by the coagulation of small a-C particles onto their surfaces. 
\end{itemize}
This diffuse ISM dust model uses $\simeq 206 - 218$\,ppm of carbon (with $\simeq 143-144$\,ppm in the a-C nano-particles), 110\,ppm of oxygen, 45\,ppm of magnesium, 32\,ppm of silicon, 19\,ppm of iron and 3\,ppm of sulphur. 
The Fe and S are incorporated within the amorphous silicate grains in the form of Fe and FeS nano-inclusions, with 7\% and 3\% volume filling fractions, respectively \citep{2013A&A...558A..62J,2014A&A...565L...9K,2015A&A...577A.110Y}.  
The breakdown of these abundances, along with the size-dependent grain material densities and the required dust masses for the above diffuse ISM model are detailed in Table~\ref{table_abundances}. The resultant nano-inclusion-containing a-Sil effective densities, $\langle \rho_{\rm a-Sil} \rangle$, are given by:
\begin{equation}
\langle \rho_{\rm a-Sil} \rangle = \rho_{\rm a-Sil} \, (1-f_{\rm Fe} - f_{\rm FeS}) + f_{\rm Fe} \, \rho_{\rm Fe} + f_{\rm FeS} \, \rho_{\rm FeS} \,, 
\label{eq_density_aSil}
\end{equation}
where $\rho_{\rm a-Sil}$, $\rho_{\rm Fe}$ and $\rho_{\rm FeS}$ are the assumed material densities (2.5, 7.87 and 4.84\,g\,cm$^{-3}$, respectively), and $f_{\rm Fe}$ (0.07) and $f_{\rm FeS}$ (0.03) are the volume filling fractions of Fe and FeS. 
The effective density of the a-C coated amorphous silicate (with nano-inclusions), and also of the large a-C:H/a-C CM grains, is then given by:
\[
\langle \rho_{\rm CM} \rangle = \frac{ m_{\rm core} + m_{\rm mantle} }{\frac{4}{3} \pi \ a^3}  
\]
\[
\ \ \ \ \ \ \ \ \ \ \ = \frac{\frac{4}{3} \pi \ \langle \rho_{\rm core} \rangle \ (a-d)^3 + \frac{4}{3} \pi \ \rho_{\rm mantle} \ [a^3-(a-d)^3]}{\frac{4}{3} \pi a^3}
\]
\begin{equation}
\ \ \ \ \ \ \ \ \ \ \ = \langle \rho_{\rm core} \rangle  \ [1-(d/a)^3]  \ +  \rho_{\rm mantle} \ \{ 1- [1-(d/a)^3] \} \,,  
\label{eq_density_aSilaC}
\end{equation}
where $a$ is the outer grain radius, $d$ is the mantle depth, $m_{\rm core}$ and $m_{\rm mantle}$ are the grain mantle and core masses and $\rho_{\rm core}$ and $\rho_{\rm mantle}$ are their  material densities. Given that in the THEMIS model the mantle thicknesses are independent of particle radius, the determination of grain densities, and consequently of the elemental abundances, requires integration over the entire size distribution for each dust component.   

We note that if the amorphous silicate Fe and FeS nano-inclusion densities are also reduced by $\simeq 20$\%, as for amorphous silicate materials, something that we have not included in our model, then the total Fe and S abundance requirements given in Table~\ref{table_abundances} would be reduced by less than 1\,ppm. 

% *********************************************************
\begin{figure*}
 \resizebox{\hsize}{!}{\includegraphics[angle=0]{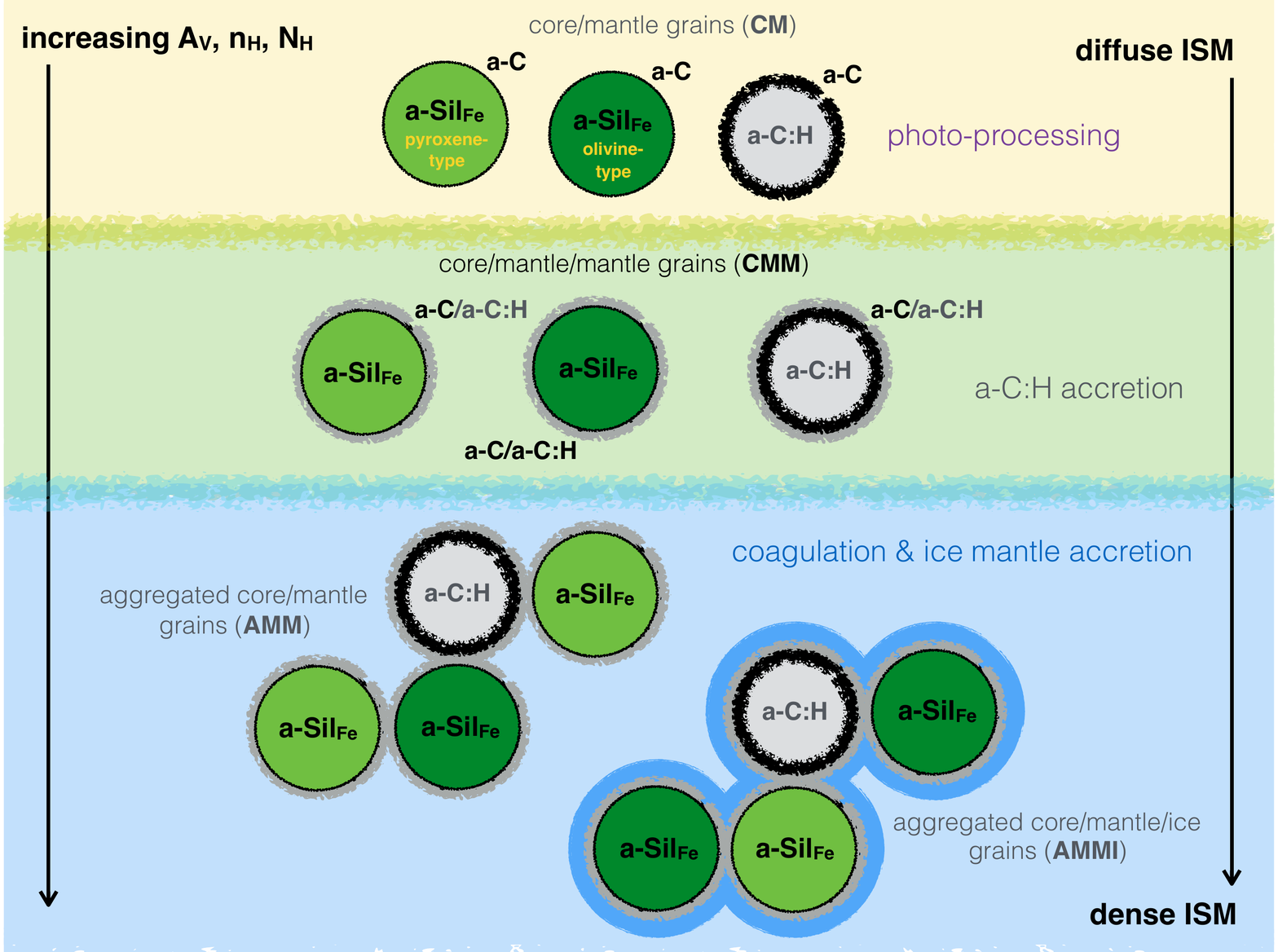}}
 \caption{A schematic view of the dust composition and stratification between diffuse ISM and dense molecular clouds. The major evolutionary processes acting on the dust in each region are indicated on the right, {\it i.e.}, photo-processing, accretion and coagulation with increasing density and extinction from top to bottom.} 
 \label{fig_grains}
\end{figure*}
% *********************************************************

%------------------------------------------------------------------
\subsection{Dust evolution within dense regions}
\label{sect_dense}
%------------------------------------------------------------------

The diffuse ISM dust model (Section~\ref{sect_model}) was extended in order to explore the evolution of the dust optical properties in the transition to denser molecular cloud regions in as self-consistent a way as possible \citep{2012A&A...548A..61K,2015A&A...579A..15K}. In THEMIS this evolution is assumed to result in the following core/mantle/mantle (CMM) grains and their aggregated forms (AMM) with ice mantles (AMMI):\footnote{We take the optical constants of ice from \cite{Warren_1984_AO}.}
\begin{itemize}
\item CMM: a-C-mantled a-Sil$_{\rm Fe,FeS}$ and a-C:H/a-C CM grains with an additional outer mantle of a-C:H formed by the accretion of remnant gas phase carbon with a composition that is determined by the local conditions (radiation field and gas density). 
\item AMM: Aggregates of CMM grains. 
\item AMMI: Aggregates of CMM grains with ice mantles. 
\end{itemize}
The accretion/coagulation-evolved dust compositions and structures, with size distributions similar to log-normal \citep{2015A&A...579A..15K}, are entirely consistent with the observed evolution of the dust properties in the transition towards denser regions \citep{2012A&A...548A..61K,Faraday_Disc_paper_2014}, including a decrease in the dust temperature and an increase in the dust spectral index and opacity at far-IR/sub-mm wavelengths \citep{2015A&A...579A..15K}. The  accretion of a-C:H mantles in denser regions is also consistent with both cloud- and core-shine (C-shine) observations \citep{2016A&A...588A..43J,2016A&A...588A..44Y}. Fig.~\ref{fig_grains} shows a schematic view of the above-described dust evolution stages and the associated grain structures and compositions. The relevant time-scales for accretion and coagulation, as a function of depth into a molecular cloud, were presented in our previous work \citep{2012A&A...548A..61K,Faraday_Disc_paper_2014} and are therefore not repeated here.

For the diffuse cloud model the dust optical properties were calculated using the Mie theory for coated spheres 
\cite[{\it e.g.},][]{1998asls.book.....B} and for more complex structures, the multi-component (CMM) and aggregate grain (AMM and AMMI) optical properties were calculated using the discrete-dipole approximation (DDA) method \citep[DDSCAT,][]{2000ascl.soft08001D}.  

The evolutionary changes in the dust structure, composition and dust size distribution, outlined above, are self-consistent and, principally due to the transformation and loss of the smallest grains through mantle accretion and coagulation onto larger grains, respectively. A full time-dependent and self-consistent study of dust accretion and coagulation in the diffuse to dense cloud transition is not yet possible. This is because the calculation of the grain optical properties, which is currently best performed with DDA, is time-consuming for complex grain structures at short wavelengths. However, these calculations now only need to be made for new grain structures. Thus, and for now, we assume a `stepped'  approach to dust evolution, {\it i.e.}, we work with a limited number of size distributions and compositions that approximate the continuity of the evolutionary processes operating in dense regions.

%------------------------------------------------------------------
\subsection{Dust evolution in PDRs and H{\footnotesize II} regions}
\label{sect_PDRs}
%------------------------------------------------------------------

In our forthcoming work we will explore the nature and the detailed physical evolution of the dust in PDRs. In these regions the dust evolution is complex and likely cannot be modelled as a simple reversal of the construction processes that lead to the assembly of complex aggregates in molecular clouds. This lack of symmetry is due to the more intense radiation fields in PDRs compared to molecular clouds, which introduces a strong hysteresis effect. 

The THEMIS approach has led to suggestions for viable routes for the formation of molecular hydrogen and of daughter hydrocarbon species in moderately excited PDRs \citep{2012A&A...540A...2J,2015A&A...581A..92J}.  
The proposed photolytic route, which has been measured in the laboratory \citep[{\it e.g.},][]{1984JAP....55..764S,1989JAP....66.3248A,1996MCP...46...198M,2011A&A...529A.146G,2014A&A...569A.119A}, could act in conjunction with and complement other formation mechanisms \citep[{\it e.g.},][]{1971ApJ...163..155H,2008ApJ...679..531R,2009ApJ...704..274L,2014A&A...569A.100B}. We have continued this work with an investigation of the effects of grain charging and the complimentary effects of gas heating, via photo-electron emission, and gas cooling 
%%%%%
(Bocchio et al. 2017, in preparation). 
%\citep{2016A&A...000A..00B}. 
The THEMIS model appears to be  consistent with the latest observations of gas heating in the diffuse ISM 
%%%%%
(Bocchio et al. 2017, in preparation). 
%\citep{2016A&A...000A..00B}. 
We are currently also exploring the likely response of the THEMIS model a-C(:H) nano-particles to hard UV photo-processing in PDRs and H{\footnotesize II} regions,  with a view to understanding nano-particle evolution in excited regions.

%------------------------------------------------------------------
\subsection{Dust processing in high energy environments}
\label{sect_highE}
%------------------------------------------------------------------

A model of the critical processing of small particles in electron and ion interactions in shocks and cosmic rays interactions was developed by \cite{2010A&A...510A..36M,2010A&A...510A..37M,2011A&A...526A..52M}.  We extended this methodology to the small a-C particles in THEMIS \citep{2012A&A...545A.124B} and used it to study the effects of electron collisional heating and dust erosion in a hot coronal gas \citep{2013A&A...556A...6B}.  We have also studied the effects of dust processing and destruction in supernova-driven shock waves using the THEMIS diffuse ISM model \citep{2014A&A...570A..32B}. 

In these environments the smallest grains appear to have rather short lifetimes and cannot be collisionally heated without also being (partly) destroyed. Recent work by \cite{Wolf:2016um} shows that the so-called super-hydrogenated polycyclic aromatic hydrocarbons,\footnote{{\it N.B.}, Small super-hydrogenated polycyclic aromatic hydrocarbons with less than 20 C atoms \citep[{\it e.g.},][]{Wolf:2016um} cannot be aromatic and should therefore strictly be called polycyclic aliphatic hydrocarbons.} which could be considered as analogues of a-C(:H) nano-particles, are less stable against photo-dissociation than their fully aromatic equivalents. Thus, and in the absence of laboratory data and models for a-C(:H) nano-particle excitation and photo-dissociative processing, the work of \cite{2010A&A...510A..36M,2010A&A...510A..37M,2011A&A...526A..52M}, which applies to fullerene and polycyclic aromatic hydrocarbons, can only be used to give an upper limit to the lifetime of a-C(:H) nano-grains with the same number of carbon atoms as fully aromatic particles.

%------------------------------------------------------------------
\subsection{Nano-particle specifics}
\label{sect_np}
%------------------------------------------------------------------

A detailed consideration of the structure of nano-particles is explicitly treated within THEMIS \citep{2012A&A...542A..98J}, where the key considerations are particle size limits on the aromatic domain distribution and particle surface passivation through hydrogenation. This approach led us to a self-consistent explanation for the origin of the UV extinction bump in a-C nano-particles \citep[$a \simeq 1$\,nm,][]{2012A&A...542A..98J} and also to viable routes to interstellar/circumstellar fullerene  formation through UV photo-processing and de-hydrogenation \citep{2012ApJ...757...41B,2012ApJ...761...35M}.

%------------------------------------------------------------------
\subsection{Comparison with observations}
\label{sect_cf_obs}
%------------------------------------------------------------------

The THEMIS dust model, including the effects of amorphous pyroxene-type silicates and Fe/FeS nano-inclusions \citep{2014A&A...565L...9K}, was recently and rather favourably compared to Planck observations of dust in the diffuse ISM \citep{2015A&A...577A.110Y,2015A&A...580A.136F}. It appears that physically-reasonable variations in the modelled grain compositions and structure (mantle thicknesses, metallic Fe inclusions, relative dust masses and size distributions) all lie within the range of the  observations \citep{2015A&A...577A.110Y}. Further, an inter-model comparison using optical and Planck data showed that the THEMIS dust model fares well and appears to have correctly predicted the observed relationship between the dust optical extinction and its long wavelength emission properties in the diffuse ISM \citep{2015A&A...577A.110Y,2015A&A...580A.136F}.  

The THEMIS dust model includes the effects of dust evolution in the denser regions of the ISM \citep{Faraday_Disc_paper_2014,2015A&A...579A..15K}. The extended model \citep{2015A&A...579A..15K} is consistent with the dust scattering properties needed to explain observations of cloud-shine and core-shine \citep[C-shine, Section~\ref{sect_dense} and][]{2016A&A...588A..43J,2016A&A...588A..44Y}. Thus, the THEMIS modelling approach encompasses, and indeed requires, rather wide variations in the composition and size distribution from region to region within the ISM, which is supported by observational studies in addition to those of \cite{2015A&A...577A.110Y} and \cite{2015A&A...580A.136F} mentioned above. Clearly, changes in the dust composition and mass can also be traced indirectly through elemental depletion studies and gas-to-dust ratio measurements in interstellar clouds. 

While the depletions of the silicate-forming elements (Si, Mg and Fe) do show rather wide variations they do follow general trends, with Si and Mg being more volatile than Fe, while Cr, Ni, Ti and Mn tend to follow Fe and show lower but seemingly well-coupled  variations \citep{2000JGR...10510257J}. These same general trends, for the same elements, are also evident in the more rigorous and complete study by \cite{2009ApJ...700.1299J}, which shows that, as the depletion strengths vary, the logarithms of the depletion factors for different elements are linearly related to one another.  What both of these studies show is that the depletions of the dust-forming elements almost always follow the same trends, {\it i.e.}, the silicate group O, Si and Mg are coupled, as are the iron group Fe, Cr, Ni, Ti and Mn, which implies that these two groups of elements are each associated with a particular interstellar dust component. Interestingly, such a division into two groups is a natural consequence of the adopted THEMIS silicate grain structure and composition. The silicate group (O, Si and Mg) represent the Mg-rich olivine- and pyroxene-type amorphous silicate matrix in the large grains with the iron group (Fe, Cr, Ni, Ti and Mn) being present as Fe-rich nano-inclusions within the amorphous silicate matrix. This segregation naturally explains why the iron group elements, embedded as nano-inclusions and protected by the surrounding silicate matrix, show weaker depletion variations with respect to the surrounding silicate group elements. This dust composition and structure is entirely consistent with the most recent experimental,  observational and direct interstellar silicate dust analysis indicators (see Section~\ref{sect_aSil}).

Significant cloud-to-cloud and within-cloud gas-to-dust variations are apparent in the low density ISM (low $n_{\rm H}$) for the same line of sight column densities \citep[$N_{\rm H}$,][]{2015ApJ...811..118R}. This work also shows a decreasing gas-to-dust ratio with decreasing dust temperature, {\it i.e.}, there is an apparent dust excess in colder regions. As \cite{2015ApJ...811..118R} state  ``\ldots grain properties may change within the clouds: they become more emissive when they are colder, while not utilizing heavy elements that already have their cosmic abundance fully locked into grains.'' 
As they noted the dust emissivity needs to be higher by about a factor of three, which is consistent with the results of \cite{2003A&A...398..551S}, \cite{2012A&A...548A..61K} and, more particularly, with those of  \cite{2015A&A...579A..15K} for the onset of dust aggregation, {\it i.e.} the CM$\rightarrow$CMM$\rightarrow$AMM dust evolution of the THEMIS model in the transition to denser cloud regions. The onset of these changes would seem to occur before ice mantles appear \citep[{\it e.g.},][]{2012ApJ...760...36P,Faraday_Disc_paper_2014} and therefore indicates that accretion onto dust and dust aggregation seemingly begins before ice mantle formation \citep[{\it e.g.},][]{2016A&A...588A..44Y}. 

Given that the dust evolution observed by \cite{2015ApJ...811..118R} cannot be driven by silicate-forming elemental depletion variations, because these elements ({\it e.g.}, Si, Mg and Fe) are maximally-depleted, this leaves the vestige carbon in the gas phase as the most likely accreting and dust emissivity-enhancing agent  \citep[{\it e.g.},][]{2013A&A...558A..62J,Faraday_Disc_paper_2014,2015A&A...577A.110Y,2016A&A...588A..43J,2016A&A...588A..44Y}. That significant dust variations are related to variations in the carbon depletion in the ISM are supported by observational  studies \citep{2012ApJ...760...36P,2015ApJ...809..120M}. 

With their measurements of interstellar carbon abundances, in both gas and dust, in environments with average hydrogen densities, $\langle n_{\rm H} \rangle$, ranging over three orders of magnitude, \cite{2012ApJ...760...36P} conclude that carbonaceous grains must be processed in the neutral ISM and that the the strength of the UV bump depends neither on the carbon abundance in dust nor on the grain-size distribution. Their work supports the suggestion that big grains form in dense clouds and that they are then progressively fragmented into smaller grains in lower density regions, which is consistent with the gradual decrease in the FUV extinction in the transition to the lower density diffuse ISM. In general agreement with the results of \cite{2012ApJ...760...36P}, \cite{2015ApJ...809..120M} find that carbon depletion into dust tends to correlate with $1/R_{\rm V}$, indicating that the FUV extinction is most likely due to small carbon grains. However, in their study they find that the strength of the UV bump does tend to correlate with the depletion of carbon into dust, in contrast to the results of \cite{2012ApJ...760...36P}. 

Recently several aspects of the fundamental nature of interstellar dust were re-examined within the framework of the THEMIS dust model, these include: the role of carbonaceous nano-particle surface chemistry and OH formation in the tenuous ISM \citep{2016RSOS....360221J}, the likely structure, form and origin of the diffuse interstellar band carriers \citep{2016RSOS....360223J} and a comprehensive re-exploration of the composition and evolution of core/mantle particles from the ISM to comets \citep{2016RSOS....360224J}. The implications arising from this wide-ranging re-appraisal of the nature of cosmic dust challenge some of our long-held views, {\it i.e.}, that it is chemically passive and essentially the same everywhere, but perhaps also open up new experimental and observational avenues that will lead to a deeper elucidation of its properties.

%------------------------------------------------------------------
\subsection{Future developments}
\label{sect_devlop}
%------------------------------------------------------------------

In our future development of THEMIS we will include appropriate, new laboratory-measured interstellar dust analogue material optical constants as and when they become available. This will be particularly important for wavelength regions where the current model data are not sufficiently constrained by such laboratory data, {\it i.e.}, for $\lambda \gtrsim 60\,\mu$m.  

In our future extensions to and applications of THEMIS we will focus on some of the key and interesting effects that nano-physics introduces to studies of interstellar nano-particles, including: photo-electron emission and gas heating, grain charge effects and the contribution of the rotational emission from a-C grains to the dust emission spectrum. 
 
In the near future we will calculate the polarising properties of all the THEMIS grain structures and use these to predict the interstellar polarisation by such grains.  

We also plan to re-visit aspects of our studies that have to date received only preliminary analysis, such as:  interactions with gas phase species \citep{Faraday_Disc_paper_2014,2015A&A...581A..92J,2016RSOS....360221J,2016RSOS....360224J}, ``volatile'' silicon in PDRs, sulphur and nitrogen depletions in the diffuse ISM \citep{2013A&A...555A..39J,2014A&A...565L...9K,2016RSOS....360221J,2016RSOS....360223J}, the origin of blue and red photoluminescence and the origin of the diffuse interstellar bands \citep{2013A&A...555A..39J,2014P&SS..100...26J,2016RSOS....360223J}.

%------------------------------------------------------------------
\section{The THEMIS model: access and links to the data} 
\label{sect_sources}
%------------------------------------------------------------------ 

Further details, developments and updates of the THEMIS model and modelling approach can be found on the THEMIS website\footnote{http://www.ias.u-psud.fr/themis/}, which will be updated as new data become available. The THEMIS model was built and developed using the interstellar dust tool DustEM\footnote{http://www.ias.u-psud.fr/dustem/}, which calculates the dust extinction, emission, {\it etc.}, when used with the appropriate input data files available via the THEMIS website. Currently the THEMIS website provides all of the input DustEM Q, C and G data files for the standard CM diffuse ISM dust model \citep{2013A&A...558A..62J,2014A&A...565L...9K}, which are also available in the current release of DustEM. 
%In addition, we provide the source complex indices of refraction data for the amorphous olivine-type and pyroxene-type silicates, without and with iron/iron sulphide nano-inclusions, and a-C(:H) particles for a selection of sizes and material band gaps. 
These will be complemented with other model data (CMM, AMM, AMMI, \ldots) as and when these data are in a fully tested and calibrated format ready for public release. Some of the more recent work on the THEMIS model has been developed within the framework of the EU FP7-funded project DustPedi\footnote{DustPedia.com} 
and THEMIS is currently the reference dust model within the DustPedia collaboration \citep{2016arXiv160906138D}. 

A number of papers have already used and made reference to the THEMIS model to explore the nature and evolution of interstellar dust in the Milky Way \citep{2015A&A...581A..92J,2016A&A...588A..43J,2016A&A...588A..44Y,2017A&A...597A.130B,2017MNRAS.465.3309D} and within other galaxies \citep{2016A&A...586A...8B,2016A&A...586A..13V,2016A&A...592A..71M,2016MNRAS.459.1646C,2016arXiv160906138D,2016arXiv161207598C}.

%------------------------------------------------------------------
%\newpage
\section{Summary}
\label{sect_summary}
%------------------------------------------------------------------

THEMIS is built around a core diffuse ISM dust model and considers the evolution of this dust in response to the physical conditions in the ambient medium. The THEMIS modelling approach is global in the sense that the observable dust properties are considered as a whole, from extreme UV to cm wavelengths, including self-consistent spectroscopic properties, and also in the sense that it can be applied to a wide variety of interstellar environments. The  modelling approach outlined here appears to provide a new view of interstellar dust and a useful framework within which many aspects of dust and its evolution in the ISM can be explored and tested. 

Recent observations have shown that the nature of interstellar dust is inherently and significantly much more complex than had previously been considered. New physically-realistic dust modelling approaches, such as that adopted in the THEMIS modelling framework, need to be stoutly anchored to the laboratory-measured properties of cosmic dust analogues, a methodology that leaves little room for arbitrary observation-fitting adjustments. The global laboratory data-constrained approach appears to be inherently more successful than models that empirically-adjust dust properties to fit the observations. While empirically-based dust models can be finely tuned to the observations and thereby advance astrophysical understanding, in achieving this they tend to sacrifice some physical understanding.

%%%%%%%%%%%%%%%%%%%%%%%%%%%%%%%%%%%%%%%%%

\begin{acknowledgements} 
The authors would like to thank their numerous colleagues for the many hours of fruitful discussions on the complex subject of interstellar dust. In particular we wish to thank Alain Abergel 
%and Marc-Antoine Miville-Desch\^enes 
for a careful reading of the manuscript. This research was, in part, made possible through the financial support of the Agence National de la Recherche (ANR) through the programs Cold Dust (ANR-07-BLAN-0364- 01) and CIMMES (ANR-11-BS56-0029) and, currently,  through the EU FP7 funded project DustPedia (Grant No. 606847).
\end{acknowledgements}

%%%%%%%%%%%%%%%%%%%%%%%%%%%%%%%%%%%%%%%%%

% for the bibliography, at the end
\bibliographystyle{aa} % style aa.bst
\bibliography{../Ant_bibliography} % your references Yourfile.bib

\end{document}